\documentclass[floats, prd, eqnum, showpacs, nofootinbib, 
twocolumn, 
eqsecnum]{revtex4-1}

\usepackage{color,graphicx}
\usepackage{amsfonts}
\usepackage{amssymb}
\usepackage{comment}

\begin{document}
    
\title{Gravitational lensing inside and outside of a marginally unstable photon sphere in a general, static, spherically symmetric, and asymptotically-flat spacetime in strong deflection limits}
\author{Naoki Tsukamoto${}^{1}$}\email{tsukamoto@rikkyo.ac.jp}
\affiliation{
${}^{1}$Department of Physics, Faculty of Science, Tokyo University of Science, 1-3, Kagurazaka, Shinjuku-ku, Tokyo 162-8601, Japan \\
}
\begin{abstract}
It is believed that rays bent inside and outside photon spheres could affect partially the black hole shadow images by the Event Horizon Telescope and the rays near photon spheres would be detected by near-future space observations.     
The investigation of the rays near the photon spheres in not only black hole spacetimes 
but also exotic spacetimes would be important since one will need them to exclude black hole mimickers.
The deflection angles of the rays deflected by the photon spheres diverge logarithmically and we can treat them by a strong-deflection-limit analysis.
The error of the strong-deflection-limit analysis becomes large if antiphoton spheres exist in the spacetimes
and the analysis breaks down when the photon spheres and the antiphoton spheres degenerate to form a marginally unstable photon sphere.
This is because the deflection angles of the rays bent by the marginally unstable photon sphere diverge in powers. 
In this paper, we extend Eiroa, Romero, and Torres's method to gravitational lensing of rays inside and outside of the marginally unstable photon sphere 
in a general, static, spherically symmetric, and asymptotically-flat spacetime in strong deflection limits
and we apply it to a Reissner-Nordstr\"{o}m spacetime and a Hayward spacetime with the marginally unstable photon sphere.
We have also confirmed that the deflection angles in the strong deflection limits by the method converge correctly to the deflection angle without approximations,  
while there are the mismatches of the coefficient of the power-divergent term of the deflection angles of the rays deflected just outside of the marginally unstable photon sphere 
in a semianalytic calculation by the author previously.
\end{abstract}

\maketitle

\section{Introduction}
Recently, the direct detection of gravitational waves by binary black holes has been reported by 
the LIGO and VIRGO Collaborations~\cite{Abbott:2016blz,LIGOScientific:2018mvr,LIGOScientific:2020ibl,KAGRA:2021vkt,LIGOScientific:2025slb} and 
the ring-shape images of supermassive black hole candidates at the centers of the giant elliptical galaxy M87 and the Milky Way Galaxy have been reported by 
Event Horizon Telescope Collaboration~\cite{Akiyama:2019cqa,EventHorizonTelescope:2022wkp}.
Strong gravity near black holes would cause high-energy phenomena 
and it would be important to detect the black holes and to understand their astrophysical environments.     

Compact objects such as black holes and wormholes have photon spheres which are spheres constructed by unstable circular light orbits due to their strong gravity~\cite{Perlick_2004_Living_Rev}.
Several aspects of the photon spheres such as the bounds of their radii~\cite{Hod:2017xkz,Peng:2018nkj,Hod:2020pim,Hod:2023jmx}, their numbers~\cite{Hod:2017zpi,Cunha:2020azh,Cunha:2024ajc,Padhye:2024yrj}, 
and their relationship to photon absorption cross sections~\cite{Sanchez:1977si,Decanini:2010fz,Wei:2011zw},
quasinormal modes~\cite{Press:1971wr,Goebel_1972,Stefanov:2010xz,Raffaelli:2014ola,Igata:2025plb,Igata:2025hpy},
a centrifugal force and gyroscopic precession~\cite{Abramowicz_Prasanna_1990,Abramowicz:1990cb,Allen:1990ci,Hasse_Perlick_2002},
Bondi's sonic horizon~\cite{Mach:2013gia,Chaverra:2015bya,Cvetic:2016bxi,Koga:2016jjq,Koga:2018ybs,Koga:2019teu},
and the stability of thin-shell wormholes~\cite{Barcelo:2000ta,Koga:2020gqd,Tsukamoto:2023kvk}, 
were investigated eagerly. 
If the gravity of compact objects is moderate, they could have also antiphoton spheres which are spheres formed by stable circular light orbits,
but the instability of the compact objects may be caused by the slow decay of linear waves near the antiphoton spheres~\cite{Keir:2014oka,Cardoso:2014sna,Cunha:2017qtt,Cunha:2017eoe,Cunha:2022gde,Zhong:2022jke}.     
The alternatives to the photon spheres and the antiphoton spheres were also suggested by several researchers~\cite{Claudel:2000yi,Koga:2019uqd,Cunha:2017eoe,Gibbons:2016isj,Shiromizu:2017ego,Yoshino:2017gqv,Galtsov:2019bty,Galtsov:2019fzq,Siino:2019vxh,Yoshino:2019dty,Cao:2019vlu,Yoshino:2019mqw,Lee:2020pre}. 

Gravitational lenses are often used to survey dark objects such as the black holes, extrasolar planets, and dark matters~\cite{Schneider_Ehlers_Falco_1992,Schneider_Kochanek_Wambsganss_2006}. 
The gravitational lenses have been mainly studied in a weak gravitational field for a hundred years since we have not had enough technology to detect astrophysical phenomena in strong gravitational fields until the last decade. 
Rays deflected by the photon spheres of the black holes and their gravitational lensing effects in strong gravitational fields have been intermittently investigated~\cite{Hagihara_1931,Darwin_1959,Atkinson_1965,Luminet_1979,Ohanian_1987,Nemiroff_1993, Frittelli_Kling_Newman_2000,Virbhadra_Ellis_2000,Bozza:2001xd,Bozza:2002zj,Perlick:2003vg,Nandi:2006ds,Virbhadra:2008ws,Hioki:2009na,Bozza_2010,Tsupko:2017rdo,Nakao:2018knn,Gralla:2019xty,Okabayashi:2020apz,Hsieh:2021scb,Aratore:2021usi,Tsupko:2022kwi,Aratore:2024bro,Feleppa:2024vdk,Feleppa:2025ejh,Tsupko:2025hhf,Saketh:2025cwf}. 
The effects of the photon spheres on the apparent shape of stars collapsing into the black holes also have been investigated~\cite{Ames_1968,Synge:1966okc,Yoshino:2019qsh,Koga:2025dii}.

It is believed that the rays deflected just inside and outside photon spheres could affect partially the observed ring images of the supermassive black hole candidates at the center of our Galaxy and the elliptic galaxy M87 by the Event Horizon Telescope~\cite{Akiyama:2019cqa,EventHorizonTelescope:2022wkp} and the rays bent near the photon spheres would be detected by near-future space observations~\cite{Lupsasca:2024xhq,Johnson:2024ttr}.   
It is important to investigate the features of the rays near the photon spheres in not only black hole spacetimes but also exotic compact objects such as wormholes~\cite{Ellis:1973yv,Chetouani:1984qdm,Perlick:2003vg,Muller:2004dq,Nandi:2006ds,Muller:2008zza,Tsukamoto:2012xs,Ohgami:2015nra,Tsukamoto:2016qro,Tsukamoto:2016zdu,Shaikh:2018kfv,Shaikh:2018oul,Bronnikov:2018nub,Shaikh:2019jfr,Bronnikov:2021liv,Maeda:2021wnl,Olmo:2023lil,Bugaev:2023mlc,Zhang:2024sgs,Xavier:2024iwr,Tsukamoto:2024pid,Bugaev:2024dte,Solodukhin:2025opw} and naked singularity~\cite{Virbhadra:1998dy,Virbhadra:2002ju,Hioki:2009na,Shaikh:2018lcc,Shaikh:2019itn} since one needs to exclude exotic compact objects or black hole mimickers to reinforce claims that the observed objects must be supermassive black holes.   

Bozza~\textit{et al.}~\cite{Bozza:2001xd} investigated
gravitational lensing of rays reflected by just outside of a photon sphere in a Schwarzschild spacetime in a strong deflection limit
$b \rightarrow b_{\mathrm{m}} + 0$, where $b$ and $b_{\mathrm{m}}$ are the impact parameter and the critical impact parameter of the ray.
The deflection angle of the rays which pass near outside of the photon sphere diverges logarithmically and it can be expressed by
\begin{eqnarray}\label{eq:defm0}
\alpha&=&-\bar{a}_+ \log \left( \frac{b}{b_{\mathrm{m}}}-1 \right) + \bar{b}_+,
\end{eqnarray}
where $\bar{a}_+$ and $\bar{b}_+$ are a coefficient and a term which are determined by the metric of the spacetime with the photon sphere.
Eiroa, Romero, and Torres suggested a method to obtain the coefficient $\bar{a}_+$ and the term $\bar{b}_+$ numerically and they applied it 
to a Reissner-Nordstr\"{o}m black hole spacetime~\cite{Eiroa:2002mk} while 
Bozza made a semianalytic formula in a general, static, spherically symmetric, and asymptotically-flat spacetime with the photon sphere
to obtain the coefficient $\bar{a}_+$ and the term $\bar{b}_+$
and got $\bar{a}_+$ in an analytic form and $\bar{b}_+$ partially numerically in the Reissner-Nordstr\"{o}m black hole spacetime~\cite{Bozza:2002zj}. 
The strong-deflection-limit analysis has been intensively studied and it has been extended over the last two decades  
~\cite{Bozza:2002af,Eiroa:2002mk,Petters:2002fa,Eiroa:2003jf,Bozza:2004kq,Bozza:2005tg,Bozza:2006sn,Bozza:2006nm,Iyer:2006cn,Bozza:2007gt,Tsukamoto:2016qro,Ishihara:2016sfv,Tsukamoto:2016oca,Tsukamoto:2016zdu,Tsukamoto:2016jzh,Shaikh:2018oul,Shaikh:2019itn,Takizawa:2021gdp,Zhang:2024sgs,Igata:2025taz}.
Tsukamoto~\cite{Tsukamoto:2016oca,Tsukamoto:2016jzh} investigated an alternative of Bozza's method~\cite{Bozza:2002zj}, 
obtained $\bar{a}_+$ and $\bar{b}_+$ in analytic forms in the Reissner-Nordstr\"{o}m black hole spacetime and confirmed that 
they are consistent with $\bar{a}_+$ and $\bar{b}_+$ obtained by Eiroa, Romero, and Torres~\cite{Eiroa:2002mk} and by Bozza~\cite{Bozza:2002zj} for the Reissner-Nordstr\"{o}m black hole spacetime.  

If we consider gravitational lensing effects of rays passing inside and outside of the photon spheres in spacetimes without event horizons~\cite{Shaikh:2019itn,Shaikh:2019jfr},    
the deflection angles of the rays in strong deflection limits~$b \rightarrow b_{\mathrm{m}} \pm 0$ can be expressed by 
\begin{eqnarray}\label{eq:def0}
\alpha&=&-\bar{a}_\pm \log \left| \frac{b}{b_{\mathrm{m}}}-1 \right| + \bar{b}_\pm,
\end{eqnarray}
where $\bar{a}_-$ and $\bar{b}_-$ are a coefficient and a term, respectively, for the ray reflected near inside of the photon sphere in the strong deflection limit $b \rightarrow b_{\mathrm{m}} - 0$ and 
they can be obtained from the metric of the spacetimes. 
Shaikh~\textit{et al.}~\cite{Shaikh:2019itn} obtained $\bar{a}_-$ as an analytical form and $\bar{b}_-$ numerically in a Reissner-Nordstr\"{o}m naked singularity spacetime
and the analytical form of $\bar{b}_-$ was obtained by Tsukamoto~\cite{Tsukamoto:2021fsz}.

The errors of deflection angles in the strong deflection limits become large if antiphoton spheres exist in the spacetimes
and the strong-deflection-limit analysis for the photon spheres breaks down 
when the photon spheres and the antiphoton spheres degenerate to be a marginally unstable photon sphere 
since the deflection angles of the rays bent by the marginally unstable photon sphere in the strong deflection limits~$b \rightarrow b_{\mathrm{m}} \pm 0$ diverge in powers.
In this paper, we treat the deflection angle of the rays in the following form:
\begin{equation}\label{eq:def01}
\alpha(b)=\frac{\bar{c}_\pm}{\left| \frac{b}{b_{\mathrm{m}}}-1\right|^\frac{1}{6} } +\bar{d}_\pm, 
\end{equation}
where $\bar{c}_+$ and $\bar{d}_+$ ($\bar{c}_-$ and $\bar{d}_-$) are a coefficient and a term, respectively, for the rays reflected by just outside (inside) of the marginally unstable photon sphere 
in the strong deflection limit~$b \rightarrow b_{\mathrm{m}} + 0$($-0$).
In Ref.~\cite{Tsukamoto:2020iez}, Tsukamoto extended Bozza's method to gravitational lensing of the rays reflected just outside of the marginally unstable photon sphere in a general, static, spherically symmetric, and asymptotically-flat spacetime in strong deflection limit~$b \rightarrow b_{\mathrm{m}} + 0$ and applied it to the Reissner-Nordstr\"{o}m naked singularity spacetime with the marginally unstable photon sphere.
Recently, Sasaki~\cite{Sasaki:2025web} investigated the details of the  gravitational lensing of the rays reflected inside and outside of the marginally unstable photon sphere in the Reissner-Nordstr\"{o}m naked singularity spacetime and obtained the higher-order terms of deflection angles in the strong deflection limits~$b \rightarrow b_{\mathrm{m}} \pm 0$.
Sasaki~\cite{Sasaki:2025web} also pointed out that the deflection angle by Tsukamoto~\cite{Tsukamoto:2020iez} in the strong deflection limit $b \rightarrow b_{\mathrm{m}} + 0$ does not converge to the deflection angle with no approximations due to the invalid value of the coefficient $\bar{c}_+$.   

In this paper, we extend Eiroa, Romero, and Torres's method~\cite{Eiroa:2002mk} to gravitational lensing of rays just inside and outside of the marginally unstable photon sphere 
in the general, static, spherically symmetric, and asymptotically-flat spacetime in strong deflection limits~$b \rightarrow b_{\mathrm{m}} \pm 0$
and we apply it to the Reissner-Nordstr\"{o}m spacetime and a Hayward spacetime~\cite{Hayward:2005gi} with the marginally unstable photon sphere.
We also confirm that mismatches in the coefficients $\bar{c}_+$ which were calculated by Tsukamoto in the semianalytic method previously~\cite{Tsukamoto:2020iez}, are modified in the new approach.     

This paper is organized as follows. 
In Sec.~II, we review the deflection angle of rays. 
In Sec.~III, we extent Eiroa, Romero, and Torres's method to cases of inside and outside of the marginally unstable photon sphere in the strong deflection limits and we apply it to the Reissner-Nordstr\"{o}m spacetime and the Hayward spacetime. In Sec.~IV, we investigate observable in gravitational lensing.
We conclude our results in Sec.~V and discuss in Sec.~VI.
We use a unit that a light speed and Newton's constant are unity in this paper.

\section{Deflection angle of rays}
We consider a general, asymptotically-flat, static and spherically symmetric spacetime with a line element, in coordinates $-\infty<t<+\infty$, $r <+\infty$, $0 \leq \vartheta \leq \pi$, and $0 \leq \varphi <2\pi$,  
\begin{equation}
\mathrm{d}s^2=-A(r)\mathrm{d}t^2+B(r)\mathrm{d}r^2+C(r)(\mathrm{d}\vartheta^2+\sin^2 \vartheta \mathrm{d}\varphi^2),
\end{equation}
with asymptotically-flat conditions  
\begin{eqnarray}\label{eq:aym_con1}
&&\lim_{r \rightarrow \infty} A(r)=\lim_{r \rightarrow \infty} B(r)=1+O(r^{-1}),\\\label{eq:aym_con2}
&&\lim_{r \rightarrow \infty} C(r)=r^2+ O(r)
\end{eqnarray}
and we assume that $A(r)$, $B(r)$, and $C(r)$ are positive and finite in a region where we consider.  
We also assume that the spacetime has circular light orbits at $r=r_\mathrm{m}$, or a condition
\begin{eqnarray}
D_\mathrm{m}=\frac{C^{\prime}_\mathrm{m}}{C_\mathrm{m}}-\frac{A^{\prime}_\mathrm{m}}{A_\mathrm{m}}=0
\end{eqnarray}
holds there.
Here, we have defined $D(r)$ as
\begin{eqnarray}
D(r)\equiv \frac{C^{\prime}(r)}{C(r)}-\frac{A^{\prime}(r)}{A(r)},
\end{eqnarray}
where $'$ denotes a differentiation with respect to $r$
and functions with the subscript m denote the functions at $r=r_\mathrm{m}$.
The circular light orbits are unstable (stable) if a condition  
\begin{eqnarray}
D_\mathrm{m}^{\prime}=\frac{C_\mathrm{m}^{\prime \prime}}{C_\mathrm{m}}-\frac{A^{\prime \prime}_\mathrm{m}}{A_\mathrm{m}}>0 \quad (<0)
\end{eqnarray}
is held and we name a sphere formed by the unstable (stable) circular light orbits a photon (an antiphoton) sphere.  
We note that the derivatives of $D(r)$ are give by
\begin{eqnarray}
D^{\prime}(r)
&=&\frac{C^{\prime \prime}(r)}{C(r)}-\frac{C^{\prime}(r)^2}{C(r)^2}\nonumber\\
&&-\frac{A^{\prime \prime}(r)}{A(r)}+\frac{A^{\prime}(r)^2}{A(r)^2}
\end{eqnarray}
and
\begin{eqnarray}
D^{\prime \prime}(r)
&=&\frac{C^{\prime \prime \prime}(r)}{C(r)}-3\frac{C^{\prime \prime}(r) C^{\prime}(r)}{C(r)^2}+2\frac{C^{\prime}(r)^3}{C(r)^3}\nonumber\\
&&-\frac{A^{\prime \prime \prime}(r)}{A(r)}+3\frac{A^{\prime \prime}(r) A^{\prime}(r)}{A(r)^2}-2\frac{A^{\prime}(r)^3}{A(r)^3}.
\end{eqnarray}

The spacetime has time-translational and axial Killing vectors $t^\mu \partial_\mu=\partial_t$ and $\varphi^\mu \partial_\mu=\partial_\varphi$
due to stationarity and axial symmetry of the spacetime, respectively,
and we can assume $\vartheta=\pi/2$ without loss of generality because of the spherical symmetry of the spacetime.

From 
\begin{eqnarray}
k^\mu k_\mu=0,
\end{eqnarray}
where $k^\mu$ is   
the wave number of a ray defined as $k^\mu \equiv \dot{x}^\mu$, where $x^\mu$ is the coordinate 
and the dot denotes a differentiation with respect to an affine parameter along the ray, 
the trajectory of the ray is expressed by
\begin{equation}\label{eq:traj_1}
-A(r)\dot{t}^2+B(r)\dot{r}^2+C(r)\dot{\varphi}^2=0.
\end{equation}
We consider that the ray, which comes from a spatial infinity, 
is reflected by a lensing object at the closest distance $r=r_0$ and goes away to the spatial infinity.
In this paper, we do not consider a case that the rays coming from the spatial infinity pass through a wormhole throat, and leave to another spatial infinity if we consider a wormhole spacetime.
At the closest distance $r=r_0$,
due to 
\begin{eqnarray}
\dot{r}_0 \equiv \left. \dot{r} \right|_{r=r_0}=0, 
\end{eqnarray}
where functions with the subscript $0$ denote the functions at the closest distance $r=r_0$,
we obtain, from Eq.~(\ref{eq:traj_1}), 
\begin{equation}\label{eq:traj_cd}
A_0\dot{t}_0^2=C_0\dot{\varphi}^2_0.
\end{equation}
From Eq.~(\ref{eq:traj_cd}), the impact parameter $b$ of the ray defined as
$b\equiv L/E$, where $E\equiv -g_{\mu \nu}t^{\mu}k^{\nu}$ and $L\equiv g_{\mu \nu} \varphi^\mu k^\nu$ 
are the conserved energy and angular momentum of the ray, respectively,
is expressed as 
\begin{equation}\label{eq:imp_2}
b=b(r_0)=\frac{L}{E}=\frac{C_0 \dot{\varphi}_0}{A_0 \dot{t}_0}= \pm \sqrt{\frac{C_0}{A_0}},
\end{equation}
and it is constant along the trajectory of the ray.
For simplicity, unless otherwise stated, we assume that the value of $b$ is positive.

By rescaling the affine parameter of the ray, the trajectory~(\ref{eq:traj_1}) is rewritten in
\begin{equation}
\dot{r}^2+V(r,b)=0,
\end{equation}
where $V(r,b)$ is an effective potential for the radial motion of the ray defined by 
\begin{equation}
V(r,b)\equiv \frac{1}{B(r)} \left( \frac{b^2}{C(r)}-\frac{1}{A(r)} \right).
\end{equation}
The ray can be in regions, where the effective potential is non-negative, i.e., $V(r,b)\leq 0$.
From Eqs.~(\ref{eq:aym_con1}) and (\ref{eq:aym_con2}),
we obtain $\lim_{r\rightarrow \infty} V(r)=-1<0$.
Thus, the ray can be at the spatial infinity.

The trajectory of the ray~(\ref{eq:traj_1}) can be rewritten in  
\begin{equation}
\left( \frac{\mathrm{d}r}{\mathrm{d}\varphi} \right)^2 
=\frac{C(r)}{B(r)}\left(\frac{C(r)}{A(r)b^2}-1\right) 
=-\frac{C(r)^2}{b^2}V(r,b) 
\end{equation}
and we obtain the deflection angle $\alpha(r_0)$ of the ray as
\begin{eqnarray}\label{eq:def_nu}
\alpha(r_0)
&\equiv& 2 \int^\infty_{r_0} \frac{\mathrm{d}r}{\sqrt{\frac{C(r)}{B(r)}\left( \frac{C(r)}{A(r)b(r_0)^2}-1 \right)}}-\pi \nonumber\\
&=& 2 \int^\infty_{r_0} \frac{b(r_0)\mathrm{d}r}{C(r)\sqrt{-V(r,b(r_0))}}-\pi.
\end{eqnarray}

\section{Deflection angle just inside and outside of the marginally unstable photon sphere in the strong deflection limits}
We extend Eiroa, Romero, and Torres's approach~\cite{Eiroa:2002mk} to the marginally unstable photon sphere in the strong deflection limits
$r_0 \rightarrow r_\mathrm{m} \pm 0$ or $b(r_0) \rightarrow b_\mathrm{m} \pm 0$,
where $b_\mathrm{m}\equiv b(r_\mathrm{m})$ is the critical impact parameter of the ray.
Here and hereinafter, the upper (lower) sign is correspond with the image of rays reflected by outside (inside) of the marginally unstable photon sphere. 

We assume that conditions 
\begin{eqnarray}
D_\mathrm{m}=\frac{C^{\prime}_\mathrm{m}}{C_\mathrm{m}}-\frac{A^{\prime}_\mathrm{m}}{A_\mathrm{m}}=0,
\end{eqnarray}
\begin{eqnarray}
D^{\prime}_\mathrm{m}=\frac{C^{\prime \prime}_\mathrm{m}}{C_\mathrm{m}}-\frac{A^{\prime \prime}_\mathrm{m}}{A_\mathrm{m}}=0,
\end{eqnarray}
and
\begin{eqnarray}
D^{\prime \prime}_\mathrm{m}=\frac{C^{\prime \prime \prime}_\mathrm{m}}{C_\mathrm{m}}-\frac{A^{\prime \prime \prime}_\mathrm{m}}{A_\mathrm{m}}>0
\end{eqnarray}
are satisfied so that the spacetime has the marginally unstable photon sphere.
Under the assumptions, we obtain 
\begin{eqnarray}
V(r_\mathrm{m},b_\mathrm{m})
=V^{\prime}(r_\mathrm{m},b_\mathrm{m})=V^{\prime \prime}(r_\mathrm{m},b_\mathrm{m})=0,
\end{eqnarray}
\begin{eqnarray}
V^{\prime \prime \prime}(r_\mathrm{m},b_\mathrm{m})
=-\frac{b_\mathrm{m}^2 D^{\prime \prime}_\mathrm{m}}{B_\mathrm{m}C_\mathrm{m}}
=-\frac{D^{\prime \prime}_\mathrm{m}}{A_\mathrm{m}B_\mathrm{m}}
<0,
\end{eqnarray}
and the deflection angles of the rays bent outside and inside of the marginally unstable photon sphere in the strong deflection limits 
$r_0 \rightarrow r_\mathrm{m} \pm 0$ or $b \rightarrow b_\mathrm{m} \pm 0$ will be expressed by   
\begin{equation}\label{eq:def5}
\alpha = \frac{c_\pm}{ \left| r_0-r_\mathrm{m} \right|^{1/2}} +\bar{d}_\pm,
\end{equation}
where $c_\pm$ are coefficients and $\bar{d}_\pm$ are finite terms. 
From the relation  
\begin{equation}\label{eq:Eiroa1}
\lim_{r_0 \rightarrow r_\mathrm{m} \pm 0}
\left[ \alpha-\frac{c_\pm}{ \left| r_0-r_\mathrm{m} \right|^{1/2}} -\bar{d}_\pm \right] =0,
\end{equation}
where 
$\alpha$ is given by Eq.~(\ref{eq:def_nu}),
and its derivative with respect to $r_0$ given by 
\begin{equation}\label{eq:Eiroa2}
\lim_{r_0 \rightarrow r_\mathrm{m} \pm 0}
\left[ \frac{\mathrm{d} \alpha}{\mathrm{d} r_0} \pm \frac{c_\pm}{ 2 \left| r_0-r_\mathrm{m} \right|^{3/2}} \right] =0,
\end{equation}
we can obtain the coefficient $c_\pm$ and the term $\bar{d}_\pm$ as, in numerical, 
\begin{equation}\label{eq:Eiroa3}
c_\pm=
\lim_{r_0 \rightarrow r_\mathrm{m} \pm 0}
\left[ \mp 2  \left| r_0-r_\mathrm{m} \right|^{3/2} \frac{\mathrm{d} \alpha}{\mathrm{d} r_0} \right] 
\end{equation}
and
\begin{eqnarray}\label{eq:Eiroa4}
\bar{d}_\pm
&=&\lim_{r_0 \rightarrow r_\mathrm{m} \pm 0}
\left[ \alpha-\frac{c_\pm}{ \left| r_0-r_\mathrm{m} \right|^{1/2}} \right] \nonumber\\
&=&\lim_{r_0 \rightarrow r_\mathrm{m} \pm 0}
\left( \alpha \pm 2 \left| r_0-r_\mathrm{m} \right| \frac{\mathrm{d} \alpha}{\mathrm{d} r_0} \right),
\end{eqnarray}
respectively.

By using series of the expansion of $b(r_0)$ around $r_0=r_\mathrm{m}$, 
\begin{equation}
b(r_0)=b_\mathrm{m}+ \frac{b_\mathrm{m} D^{\prime \prime}_\mathrm{m}}{12} \left( r_0 -r_\mathrm{m} \right)^3 + O\left( \left( r_0 -r_\mathrm{m} \right)^4 \right),
\end{equation}
the deflection angle~(\ref{eq:def5}) can be expressed by
\begin{equation}
\alpha(b)= \frac{\bar{c}_\pm}{\left| \frac{b}{b_\mathrm{m}}-1 \right|^{1/6} } +\bar{d}_\pm,
\end{equation}
where $\bar{c}_\pm$ is given by 
\begin{equation}\label{eq:Eiroa10}
\bar{c}_\pm \equiv \left| \frac{D^{\prime \prime}_\mathrm{m}}{12} \right|^{1/6} c_\pm.
\end{equation}

\subsection{Reissner-Nordstr\"{o}m spacetime}
In the Reissner-Nordstr\"{o}m black hole spacetime,
gravitational lensing of rays bent by outside of a photon sphere in the strong deflection limit $b \rightarrow b_\mathrm{m} + 0$ was investigated by 
Eiroa, Romero, and Torres~\cite{Eiroa:2002mk} in a numerical way and by Bozza~\cite{Bozza:2002zj} in a semianalytical way, and 
extended source effects~\cite{Eiroa:2002mk}, the time delay of rays~\cite{Sereno:2003nd}, retrolensing~\cite{Eiroa:2003jf,Tsukamoto:2016oca,Tsukamoto:2016jzh}, and
higher-order terms of the deflection angle~\cite{Tsukamoto:2022tmm,Tsukamoto:2022uoz,Sasaki:2023rdf} were studied.
The images of light sources at near the photon sphere in the Reissner-Nordstr\"{o}m black hole spacetime were investigated in Ref.~\cite{Aratore:2024bro,Tsupko:2025hhf} 
by using a strong-deflection-limit analysis with arbitrary source distances~\cite{Bozza:2007gt}.
In the Reissner-Nordstr\"{o}m naked singularity spacetime, the size of the photon sphere or the shadows~\cite{Zakharov:2014lqa}
and gravitational lensing of rays deflected inside and outside of a photon sphere in the strong deflection limits $b \rightarrow b_\mathrm{m} \pm 0$ were investigated in Refs.~\cite{Shaikh:2019itn,Tsukamoto:2021fsz,Tsukamoto:2021lpm}.

Tsukamoto~\cite{Tsukamoto:2020iez} got the deflection angle
with the coefficient 
\begin{equation}\label{eq:cRN3}
\bar{c}_+=2^{5/3}3^{1/2} \sim 5.49892
\end{equation}
and the constant 
\begin{equation}\label{eq:dRN3}
\bar{d}_+= -\sqrt{6} -\pi \sim -5.59108
\end{equation}
in the strong deflection limit~$b \rightarrow b_{\mathrm{m}} + 0$ in the semianalytical method.
Sasaki~\cite{Sasaki:2025web} investigated the deflection angle of the rays reflected inside and outside of the marginally unstable photon sphere in the Reissner-Nordstr\"{o}m naked singularity spacetime 
in the strong deflection limits~$b \rightarrow b_{\mathrm{m}} \pm 0$ and obtained 
\begin{equation}\label{eq:cRN2}
\bar{c}_+=\frac{\Gamma \left(\frac{1}{12}\right) \Gamma \left(\frac{7}{12}\right)}{2^{1/6} \sqrt{3} \Gamma \left(\frac{2}{3}\right)} \sim 6.67748,
\end{equation}
\begin{equation}\label{eq:dRN2}
\bar{d}_+= -\sqrt{6} -\pi \sim -5.59108,
\end{equation}
\begin{equation}\label{eq:cRN5}
\bar{c}_- =\sqrt{3}\bar{c}_+  \sim 11.5657,
\end{equation}
and
\begin{equation}\label{eq:dRN5}
\bar{d}_-= \bar{d}_+= -\sqrt{6} -\pi \sim  -5.59108
\end{equation}
where $\Gamma \left( z \right)$ is the gamma function,
and higher terms in the deflection angles.    
Sasaki pointed out that $\bar{c}_+$ in Eq.~(\ref{eq:cRN3}) is invalid
since that the deflection angle with $\bar{c}_+$ in Eq.~(\ref{eq:cRN3}) and $\bar{d}_+$ in Eq.~(\ref{eq:dRN3}) obtained by Tsukamoto~\cite{Tsukamoto:2020iez} does not converge to the deflection angle without approximations.   

The Reissner-Nordstr\"{o}m spacetime 
with the functions of the metric  
\begin{eqnarray}
&&A(r)=B(r)^{-1}=1-\frac{2M}{r}+\frac{9M^2}{8r^{2}},\\
&&C(r)=r^{2},
\end{eqnarray} 
has the marginally unstable photon sphere,  
which is formed by rays with the critical impact parameter
\begin{equation}
b_{\mathrm{m}}=\frac{3\sqrt{6}M}{2}
\end{equation}
at $r=r_{\mathrm{m}}$, 
where $r_{\mathrm{m}}$ is given by 
\begin{equation}
r_{\mathrm{m}}=\frac{3M}{2},
\end{equation}
and it holds the conditions 
\begin{eqnarray}
&&D_\mathrm{m}=D^{\prime}_\mathrm{m}=0, \\
&&D^{\prime \prime}_\mathrm{m}=\frac{64}{9}>0.
\end{eqnarray}
Figure~1 shows the effective potential $V(r/M,b)$ for the radial motion of the rays.
\begin{figure}[htbp]
\begin{center}
\includegraphics[width=87mm]{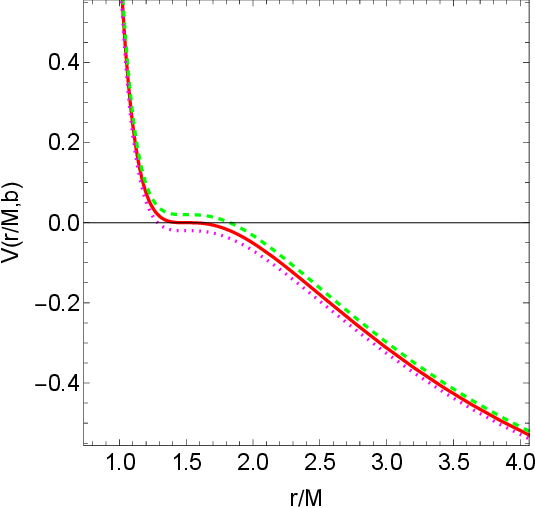}
\end{center}
\caption{The effective potential $V(r/M,b)$ in the Reissner-Nordstr\"{o}m spacetime with the marginally unstable photon sphere is shown.
The broken~(green), solid~(red), and dotted~(magenta) curves denote $V(r/M,b)$ for $b=0.99 b_{\mathrm{m}}$, $b_{\mathrm{m}}$, and $1.01 b_{\mathrm{m}}$, respectively.}
\end{figure}

From Eqs.~(\ref{eq:Eiroa3}), (\ref{eq:Eiroa4}), and (\ref{eq:Eiroa10}), we obtain numerically
\begin{equation}\label{eq:cRN1}
\bar{c}_+ \sim 6.67748
\end{equation}
and
\begin{equation}\label{eq:dRN1}
\bar{d}_+ \sim -5.59011
\end{equation}
for the rays just outside of the marginally unstable photon sphere, and 
\begin{equation}\label{eq:cRN4}
\bar{c}_- \sim 11.5658
\end{equation}
and
\begin{equation}\label{eq:dRN4}
\bar{d}_- \sim -5.64071
\end{equation}
for the ones inside of it 
in the Reissner-Nordstr\"{o}m spacetime.

The deflection angles $\alpha$ are shown in Fig.~2
and the percent error of the deflection angle of Eq.~(3.12) defined by
\begin{equation}
\frac{\alpha \:  \mathrm{of \: Eq.} (3.12)-\alpha \: \mathrm{of \: Eq.} (2.17)}{\alpha \: \mathrm{of \: Eq.} (2.17)} \times 100
\end{equation}
is shown in Fig.~3.
From Figs.~2 and 3, we confirm that our results are the almost same as ones by Sasaki~\cite{Sasaki:2025web} and that they converge to the deflection angle~(2.17) 
in the strong deflection limits $b \rightarrow b_\mathrm{m} \pm 0$. 
At a glance at Fig.~2, it might seem as though the deflection angle by Tsukamoto~\cite{Tsukamoto:2020iez} converge to the deflection angle~(2.17) in the strong deflection limit $b \rightarrow b_\mathrm{m} + 0$
but Fig. 3 shows that it does not converge as pointed out by Sasaki~\cite{Sasaki:2025web}.
\begin{figure}[htbp]
\begin{center}
\includegraphics[width=87mm]{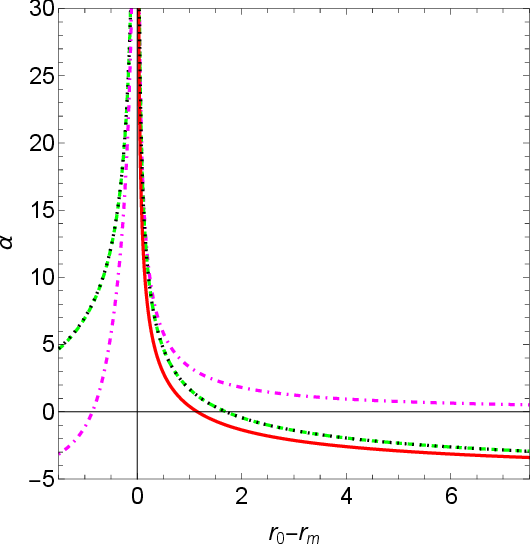}
\end{center}
\caption{
The deflection angles~$\alpha$ in the Reissner-Nordstr\"{o}m spacetime by using
Eq.~(2.17) is denoted by the dot-dashed~(magenta) curve.
The one by using 
Eq.~(3.12) with $\bar{c}_+$ of Eq.~(\ref{eq:cRN1}), $\bar{d}_+$ of Eq.~(\ref{eq:dRN1}), $\bar{c}_-$ of Eq.~(\ref{eq:cRN4}), and $\bar{d}_-$ of Eq.~(\ref{eq:dRN4}) in our numerical method,
with $\bar{c}_+$ of Eq.~(\ref{eq:cRN2}), $\bar{d}_+$ of Eq.~(\ref{eq:dRN2}), $\bar{c}_-$ of Eq.~(\ref{eq:cRN5}), and $\bar{d}_-$ of Eq.~(\ref{eq:dRN5}) by Sasaki~\cite{Sasaki:2025web}, 
and with $\bar{c}_+$ of Eq.~(\ref{eq:cRN3}) and $\bar{d}_+$ of Eq.~(\ref{eq:dRN3}) by Tsukamoto~\cite{Tsukamoto:2020iez}
are denoted by the dashed~(green), dotted~(black), and solid~(red) curves, respectively.
Notice that the dashed~(green) and dotted~(black) curves are overlapped. 
}
\end{figure}
\begin{figure}[htbp]
\begin{center}
\includegraphics[width=87mm]{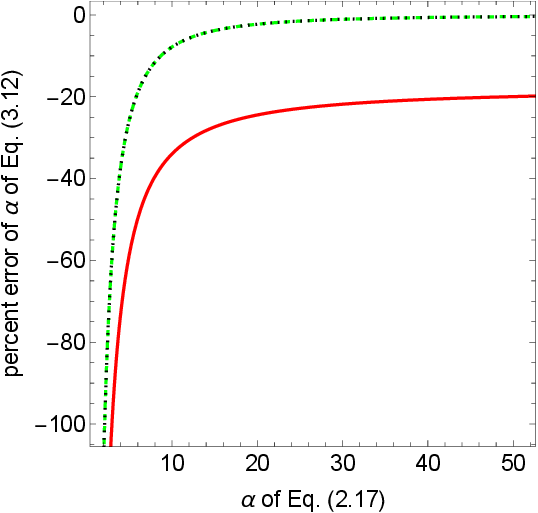}
\includegraphics[width=87mm]{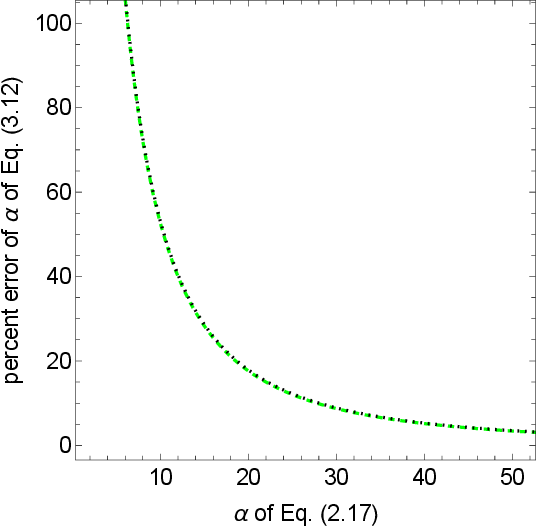}
\end{center}
\caption{The percent errors of deflection angles~$\alpha$ of Eq.~(3.12) against the deflection angle~$\alpha$ of Eq.~(2.17) 
in the Reissner-Nordstr\"{o}m spacetime are plotted in top and bottom panels for rays producing images outside and inside of the marginally unstable photon sphere, respectively. 
In the top panel, dashed~(green), dotted~(black), solid~(red) curves denote the percent errors of deflection angles~$\alpha$ of Eq.~(3.12) 
with $\bar{c}_+$ of Eq.~(\ref{eq:cRN1}) and $\bar{d}_+$ of Eq.~(\ref{eq:dRN1}) in our numerical method, 
with $\bar{c}_+$ of Eq.~(\ref{eq:cRN2}) and $\bar{d}_+$ of Eq.~(\ref{eq:dRN2}) by Sasaki~\cite{Sasaki:2025web}, and   
with $\bar{c}_+$ of Eq.~(\ref{eq:cRN3}) and $\bar{d}_+$ of Eq.~(\ref{eq:dRN3}) by Tsukamoto~\cite{Tsukamoto:2020iez}, respectively. 
In the bottom panel, dashed~(green) and dotted~(black) curves denote the percent errors of deflection angles~$\alpha$ of Eq.~(3.12) 
with $\bar{c}_-$ of Eq.~(\ref{eq:cRN4}) and $\bar{d}_-$ of Eq.~(\ref{eq:dRN4}) in our numerical method, and 
with $\bar{c}_-$ of Eq.~(\ref{eq:cRN5}) and $\bar{d}_-$ of Eq.~(\ref{eq:dRN5}) by Sasaki~\cite{Sasaki:2025web}, respectively. 
Note that the dashed~(green) and dotted~(black) curves are overlapped. 
}
\end{figure}

\subsection{Hayward spacetime}
The shadow images of the photon sphere of Hayward black holes~\cite{Li:2013jra,Tsukamoto:2017fxq} and the deflection angle of a light in a weak-field approximation 
and in the strong deflection limit $b \rightarrow b_{\mathrm{m}}+0$ in the Hayward black hole spacetime~\cite{Wei:2015qca} were studied. 
Chiba and Kimura~\cite{Chiba:2017nml} showed the shadow images in black hole and overcharged cases.

Chiba and Kimura~\cite{Chiba:2017nml} investigated the deflection angle of the rays reflected by the marginally unstable photon sphere and obtained the coefficient $\bar{c}_+$ as,
from Eqs.~(21) and (22) in Ref.~\cite{Chiba:2017nml}, 
\begin{equation}\label{eq:cH2}
\bar{c}_+ =\frac{c_+ M^{1/6}}{b_\mathrm{m}^{1/6} } \sim 6.01316, 
\end{equation}
where 
\begin{equation}
c_+=2^{11/6}3^{2/3}5^{5/4} \int^\infty_0\frac{\mathrm{d}y}{\sqrt{432y+900y^2+625y^3}} \sim 7.771,
\end{equation}
but they do not consider the constants $\bar{d}_+$ and $\bar{d}_-$ and the coefficient~$\bar{c}_-$.   
Tsukamoto~\cite{Tsukamoto:2020iez} studied the deflection angle of the rays reflected outside of the marginally unstable photon sphere in the strong deflection limit~$b \rightarrow b_\mathrm{m} + 0$ 
and got the coefficient as, by using the semianalytic method, 
\begin{eqnarray}\label{eq:cH3}
\bar{c}_+ \sim \frac{2^{13/6} 3^{1/3}}{5^{1/6}} \sim 4.95196
\end{eqnarray}
and the constant
\begin{equation}\label{eq:dH3}
\bar{d}_+= -4\sqrt{\frac{6}{5}}+I_{\mathrm{R}}(r_{\mathrm{m}}) -\pi \sim -5.62607,
\end{equation}
where an integral
\begin{equation}\label{eq:dH3}
I_{\mathrm{R}}(r_{\mathrm{m}}) = 2 \int^1_0  \left[ \sqrt{\frac{6-3z+3z^2-z^3}{z^3(5-5z+z^2)}}- \sqrt{\frac{6}{5z^3}} \right] \mathrm{d}z
\end{equation}
was calculated in numerical.

The Hayward spacetime~\cite{Hayward:2005gi} has the marginally unstable photon sphere at $r=r_{\mathrm{m}}=25M/12$ made by the rays with 
the critical impact parameter 
\begin{equation}
b_{\mathrm{m}}=\frac{25\sqrt{5}M}{12}
\end{equation}
and it holds the conditions 
\begin{eqnarray}
&&D_\mathrm{m}=D^{\prime}_\mathrm{m}=0, \\
&&D^{\prime \prime}_\mathrm{m}=\frac{1728}{625}>0
\end{eqnarray}
when the functions of the metric are
\begin{eqnarray}
&&A(r)=B(r)^{-1}=1-\frac{2Mr^2}{r^3+2q^2M},\\
&&C(r)=r^{2},
\end{eqnarray}
where we have set $2q^2 M=5^5 M^3/(2^6 3^3)$.
The effective potentials~$V(r/M,b)$ of the rays for the radial motion are plotted in Fig.~4.
\begin{figure}[htbp]
\begin{center}
\includegraphics[width=87mm]{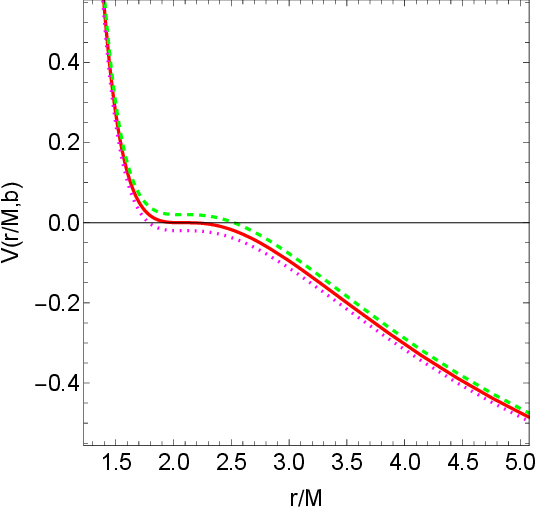}
\end{center}
\caption{The effective potentials $V(r/M,b)$ for $b=0.99 b_{\mathrm{m}}$, $b_{\mathrm{m}}$, and $1.01 b_{\mathrm{m}}$ in the Hayward spacetime with the marginally unstable photon sphere 
are shown as the broken~(green), solid~(red), and dotted~(magenta) curves, respectively.}
\end{figure}
From Eqs.~(\ref{eq:Eiroa3}), (\ref{eq:Eiroa4}), and (\ref{eq:Eiroa10}), we obtain in numerical 
\begin{equation}\label{eq:cH1}
\bar{c}_+  \sim 6.01324
\end{equation}
and
\begin{equation}\label{eq:dH1}
\bar{d}_+  \sim -5.60958
\end{equation}
for the rays outside of the marginally unstable photon sphere
%
\begin{equation}\label{eq:cH4}
\bar{c}_-  \sim  10.4154
\end{equation}
and
\begin{equation}\label{eq:dH4}
\bar{d}_-  \sim  -5.67374
\end{equation}
for the ones inside of the marginally unstable photon sphere in the Hayward spacetime.

The deflection angles $\alpha$ in the Hayward spacetime with the marginally unstable photon sphere are shown in Fig.~5
and the percent errors of the deflection angles of Eq.~(3.12) with the coefficients~$\bar{c}_\pm$ and the constants~$\bar{d}_\pm$ are shown in Fig.~6.
From Fig.~5, one may consider that the deflection angle by Tsukamoto~\cite{Tsukamoto:2020iez} converges to the deflection angle~(2.17) in the strong deflection limit $b \rightarrow b_\mathrm{m} + 0$
but Fig.~6 shows that it does not converge.
From Fig.~6, we find that the deflection angle which we obtained in the strong deflection limits $b \rightarrow b_\mathrm{m} \pm 0$ 
and one by Chiba and Kimura~\cite{Chiba:2017nml} in the strong deflection limit $b \rightarrow b_\mathrm{m} + 0$ converges to the deflection angle~(2.17)
and we notice that the errors of our calculations are smaller than one by Chiba and Kimura~\cite{Chiba:2017nml} due to their lack of the constant $\bar{d}_+$.   
Therefore, we recognize that the coefficient $\bar{c}_+$~(\ref{eq:cH2}) by Chiba and Kimura~\cite{Chiba:2017nml} and the constant $\bar{d}_+$~(\ref{eq:dH3}) by Tsukamoto~\cite{Tsukamoto:2020iez}
are correct while the coefficient $\bar{c}_+$~(\ref{eq:cH3}) by using the semianalytic method in Ref.\cite{Tsukamoto:2020iez} should be modified.
\begin{figure}[htbp]
\begin{center}
\includegraphics[width=87mm]{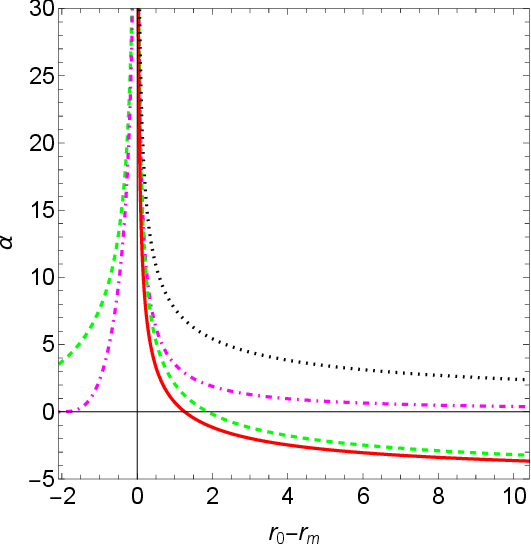}
\end{center}
\caption{The deflection angles~$\alpha$ in the Hayward spacetime with the marginally unstable photon sphere are shown.
The dot-dashed~(magenta) curve denotes the deflection angle~$\alpha$ of Eq.~(2.17).
The dashed~(green), dotted~(black), and solid~(red) 
curves denote the deflection angles~$\alpha$ of Eq.~(3.12) 
with $\bar{c}_+$ of Eq.~(\ref{eq:cH1}), $\bar{d}_+$ of Eq.~(\ref{eq:dH1}), $\bar{c}_-$ of Eq.~(\ref{eq:cH4}), and $\bar{d}_-$ of Eq.~(\ref{eq:dH4}) in our numerical method,
with $\bar{c}_+$ of Eq.~(\ref{eq:cH2}) and $\bar{d}_+=0$  by Chiba and Kimura~\cite{Chiba:2017nml}, and   
with $\bar{c}_+$ of Eq.~(\ref{eq:cH3}) and $\bar{d}_+$ of Eq.~(\ref{eq:dH3}) by Tsukamoto~\cite{Tsukamoto:2020iez}, respectively.
}
\end{figure}
\begin{figure}[htbp]
\begin{center}
\includegraphics[width=87mm]{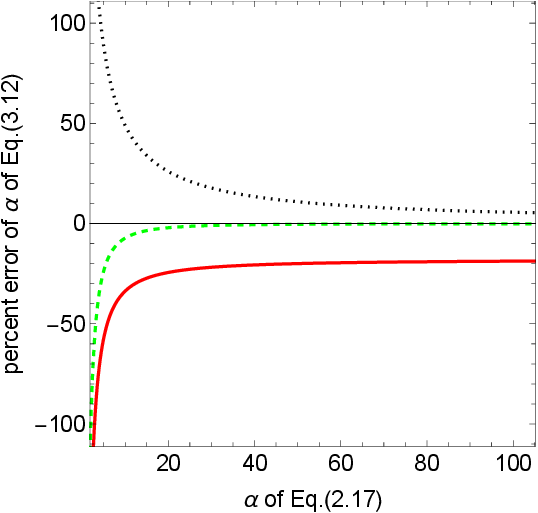}
\includegraphics[width=87mm]{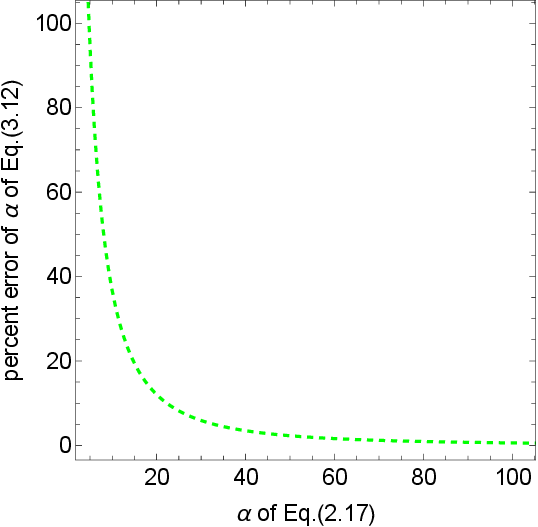}
\end{center}
\caption{The percent errors of the deflection angles~$\alpha$ of Eq.~(3.12) against the deflection angle~$\alpha$ of Eq.~(2.17) 
in the Hayward spacetime are plotted in top and bottom panels for the rays producing images outside and inside of the marginally unstable photon sphere, respectively.
In the top panel, the dashed~(green), dotted~(black), and solid~(red) curves denote the percent errors of the deflection angles~$\alpha$ of Eq.~(3.12) 
with $\bar{c}_+$ of Eq.~(\ref{eq:cH1}) and $\bar{d}_+$ of Eq.~(\ref{eq:dH1}) in our numerical method, 
with $\bar{c}_+$ of Eq.~(\ref{eq:cH2}) and $\bar{d}_+=0$ by Chiba and Kimura~\cite{Chiba:2017nml}, and  
with $\bar{c}_+$ of Eq.~(\ref{eq:cH3}) and $\bar{d}_+$ of Eq.~(\ref{eq:dH3}) by Tsukamoto~\cite{Tsukamoto:2020iez}, respectively.
In the bottom panel, the dashed~(green) curve denote the percent error of the deflection angle~$\alpha$ of Eq.~(3.12) 
with $\bar{c}_-$ of Eq.~(\ref{eq:cH4}), and $\bar{d}_-$ of Eq.~(\ref{eq:dH4}) in our numerical method.
}
\end{figure}

\section{Observable of gravitational lensing in the strong deflection limits}
In this section, we consider the observable of gravitational lensing of the rays bent inside and outside of the marginally unstable photon sphere 
in the strong deflection limits $b \rightarrow b_{\mathrm{m}} \pm 0$ or $r \rightarrow r_{\mathrm{m}} \pm 0$.
We consider that a light source S with a source angle $\phi$ emits the ray with an impact parameter $b$, 
a lens L reflects the ray by a deflection angle $\alpha$ 
and an observer O sees the source S as an image I with an image angle $\theta$. 
The lens configuration is shown as Fig.~7.

\begin{figure}[htbp]
\begin{center}
\includegraphics[width=80mm]{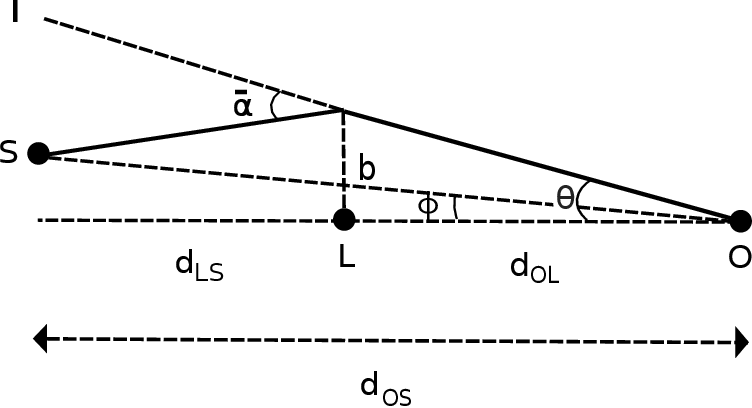}
\end{center}
\caption{A small lens configuration with an effective deflection angle $\left| \bar{\alpha} \right| \ll 1$, an image angle $\left| \theta \right| \ll 1$, and a source angle $\left| \phi \right| \ll 1$.
A ray emitted by a light source S with an impact parameter $b$ is bent by a lens object L and it reaches to an observer O who observes an image I.   
$d_{\mathrm{OS}}$, $d_{\mathrm{LS}}$, and $d_{\mathrm{OL}}$ are angular distances between O and S, between L and S, and between O and L, respectively.}
\end{figure}
We introduce an effective deflection angle $\bar{\alpha}$ of the ray defined by
\begin{eqnarray}
\bar{\alpha} \equiv \alpha \quad \mathrm{mod} \;  2 \pi
\end{eqnarray}
and the deflection angle $\alpha$ can be expressed by 
\begin{eqnarray}\label{eq:baralpha}
\alpha=\bar{\alpha}+ 2 \pi n,
\end{eqnarray}
where a non-negative integer $n$ is the winding number of the ray. 

We assume that the angles $\left| \bar{\alpha} \right|$, $\left| \theta \right|$, and $\left| \phi \right|$ are small, i.e., $\left| \bar{\alpha} \right|, \left| \theta \right|, \left| \phi \right| \ll 1$.
Then, a lens equation \cite{Bozza:2008ev} is obtained as
\begin{eqnarray}\label{eq:lens}
\theta=\frac{d_{\mathrm{LS}}}{d_{\mathrm{OS}}}\bar{\alpha}+\phi,
\end{eqnarray}
where $d_{\mathrm{LS}}$ and $d_{\mathrm{OS}}$ are angular distances 
between L and S and between O and S, respectively,
and the image angle $\theta$ can be expressed by 
\begin{eqnarray}
\theta= \frac{b}{d_{\mathrm{OL}}},
\end{eqnarray}
where $d_{\mathrm{OL}}=d_{\mathrm{OS}}-d_{\mathrm{LS}}$ is an angular distance between O and L.

We express the deflection angle $\alpha$ of the ray reflected just inside and outside of the marginally unstable photon sphere and its derivative respective to the image angle $\theta$ as
\begin{equation}\label{eq:alpha}
\alpha(\theta)=\frac{\bar{c}_\pm}{\left| \frac{\theta}{\theta_\infty}-1 \right|^{1/6}} +\bar{d}_\pm, 
\end{equation}
and 
\begin{equation}\label{eq:dalphadtheta}
\frac{d\alpha}{d\theta}=\mp \frac{\bar{c}_\pm}{ 6 \theta_\infty  \left| \frac{\theta}{\theta_\infty}-1 \right|^{7/6}}, 
\end{equation}
respectively,
where $\theta_\infty$ defined by 
\begin{equation}
\theta_\infty \equiv \frac{b_{\mathrm{m}}}{d_{\mathrm{OL}}}
\end{equation}
is an image angle for the marginally unstable photon sphere. 
By substituting $\theta^0_{n\pm}$, which is defined by 
\begin{eqnarray}\label{eq:theta0n}
\alpha \left(\theta^0_{n\pm} \right)=2\pi n
\end{eqnarray}
for each positive winding number $n\geq 1$,
into the deflection angle $\alpha(\theta)$ expressed by Eq.~(\ref{eq:alpha}) of the ray reflected just inside and outside of the marginally unstable photon sphere, 
we obtain $\theta^0_{n\pm}$ as 
\begin{eqnarray}\label{eq:theta0n2}
\theta^0_{n\pm} = \left[ 1 \pm \left( \frac{\bar{c}_\pm}{2\pi n - \bar{d}_\pm} \right)^6 \right] \theta_\infty.
\end{eqnarray}
We expand the deflection angle $\alpha\left(\theta \right)$ around $\theta=\theta^0_{n\pm}$ for each positive winding number $n \geq 1$ as 
\begin{eqnarray}\label{eq:alphaexpand}
\alpha(\theta) 
&=&\alpha \left(\theta^0_{n\pm} \right)+\left.\frac{\mathrm{d}\alpha}{\mathrm{d}\theta}\right|_{\theta=\theta^0_{n\pm}} \left(\theta-\theta^0_{n\pm} \right) \nonumber\\
&&+O\left( \left( \theta-\theta^0_{n\pm} \right)^2 \right). 
\end{eqnarray} 
From Eqs. (\ref{eq:baralpha}),  (\ref{eq:dalphadtheta}), (\ref{eq:theta0n}), and (\ref{eq:alphaexpand}), 
the effective deflection angle $\bar{\alpha}\left(\theta_{n\pm} \right)$, where $\theta=\theta_{n\pm}$ is the image angle for each positive winding number $n \geq 1$, is given by
\begin{eqnarray}\label{eq:baralpha2}
\bar{\alpha}\left(\theta_{n\pm} \right)=\mp \frac{\bar{c}_\pm \left( \theta_{n\pm} -\theta^0_{n\pm} \right) }{6 \theta_\infty \left| \frac{\theta^0_{n\pm}}{\theta_\infty} -1 \right|^{7/6}}.
\end{eqnarray}

From Eqs.~(\ref{eq:lens}),  (\ref{eq:theta0n2}), and (\ref{eq:baralpha2}),
the image angles $\theta_{n\pm}=\theta_{n\pm}(\phi)$  of the rays reflected inside and outside of the marginally unstable photon sphere for each positive winding number $n \geq 1$ are obtained as  
\begin{equation}
\theta_{n\pm}(\phi)\sim \theta^0_{n\pm} \mp \frac{d_{\mathrm{OS}}}{d_{\mathrm{LS}}}   \frac{6 \bar{c}_\pm^6}{ \left( 2 \pi n - \bar{d}_\pm \right)^{7}} \theta_\infty (\theta^0_{n\pm} -\phi).
\end{equation}
If the observer O, the lens L, and the source S are aligned in a straight line in this order, the observer sees ring-shaped images called Einstein rings.  
By setting $\phi=0$, we obtain the image angle of the Einstein ring for each positive winding number $n \geq 1$ as
\begin{equation}
\theta_{\mathrm{E}n\pm} \equiv \theta_{n\pm}(0)\sim \theta^0_{n\pm} \left[1 \mp \frac{d_{\mathrm{OS}}}{d_{\mathrm{LS}}}   \frac{6 \bar{c}_\pm^6 }{ \left( 2 \pi n - \bar{d}_\pm \right)^{7}} \theta_\infty \right].
\end{equation}
The difference of the image angles between the image with $n=1$ and the marginally unstable photon sphere with $n=\infty$
are obtained as 
\begin{equation}
\bar{s}_\pm \equiv \theta_{1\pm} - \theta_\infty \sim \theta^0_{1\pm} - \theta^0_\infty = \pm \left( \frac{\bar{c}_\pm}{2\pi - \bar{d}_\pm} \right)^6 \theta_\infty.
\end{equation}

The magnifications of the images of the ray reflected by just inside and outside of the marginally unstable photon sphere for each positive winding number $n \geq 1$ are given by
\begin{eqnarray}
\mu_{n\pm}(\phi) 
&\equiv& \frac{\theta_{n\pm}}{\phi}\frac{\mathrm{d}\theta_{n\pm}}{\mathrm{d}\phi} \nonumber\\
&\sim& \frac{d_{\mathrm{OS}}}{d_{\mathrm{LS}}}  \frac{6 \bar{c}_\pm^6 \left[ \bar{c}_\pm^6 \pm (2\pi n-\bar{d}_\pm)^6 \right]}{(2\pi n-\bar{d}_\pm)^{13}} \frac{\theta_\infty^2}{\phi}.
\end{eqnarray}

Every gravitational lensed image with the winding number $n \geq 1$ and with a positive impact parameter makes a pair of the image with the same winding number $n \geq 1$ and with a negative impact parameter.
The pair of images with every winding number $n \geq 1$ has the almost same absolute values of the image angles and the magnifications but a different signs under the assumptions of the small angles.
Thus, the diameter of the paired images is given by $2 \theta_{n \pm }$ with the winding number $n \geq 1$ for the positive impact parameter 
and the total magnification $\mu_{n \mathrm{tot} \pm}$, which is defined by the sum of the absolute values of the magnification of the pair of the images with the winding number $n \geq 1$, is given by       
\begin{eqnarray}
\mu_{n \mathrm{tot} \pm} \sim 2 \left| \mu_{n \pm} \right|. 
\end{eqnarray}
  
The ratio of the magnification of the image with $n=1$ to the sum of the other images with $n\geq 2$ is given by 
\begin{eqnarray}
\bar{r}_\pm \equiv \frac{\mu_{1\pm}}{\sum^\infty_{n=2} \mu_{n\pm}} \sim \frac{K_{1\pm}}{\sum^\infty_{n=2} K_{n\pm}},
\end{eqnarray}
where $K_{n\pm}$ for each positive winding number $n \geq 1$ is defined by  
\begin{eqnarray}
K_{n\pm}\equiv \frac{\bar{c}_\pm^6 \pm (2\pi n-\bar{d}_\pm)^6}{(2\pi n-\bar{d}_\pm)^{13}}.
\end{eqnarray}

\section{Conclusion}

In this paper, we have extent Eiroa, Romero, and Torres's method~\cite{Eiroa:2002mk} to gravitational lensing of rays just inside and outside of a marginally unstable photon sphere 
in a general, static, spherically symmetric, and asymptotically-flat spacetime in the strong deflection limits~$b \rightarrow b_{\mathrm{m}} \pm 0$
and we have applied it to a Reissner-Nordstr\"{o}m spacetime and a Hayward spacetime with 
the marginally unstable photon sphere.

We have confirmed that the deflection angles in the strong deflection limits~$b \rightarrow b_{\mathrm{m}} \pm 0$ by the method converge correctly to the deflection angle without approximations,  
while there were mismatches in semianalytic calculations by the author previously~\cite{Tsukamoto:2020iez}, which was pointed out by Sasaki~\cite{Sasaki:2025web}, 
for the coefficient $\bar{c}_+$ of the power-divergent term of the deflection angles of the rays deflected just outside of the marginally unstable photon sphere in the Reissner-Nordstr\"{o}m spacetime.  

In Ref.~\cite{Sasaki:2025web}, Sasaki found that the term $\bar{d}_-$ is the same as the term $\bar{d}_+$ analytically in the Reissner-Nordstr\"{o}m spacetime with the marginally unstable photon sphere.
We obtained $\bar{d}_+ \sim -5.59011$ and $\bar{d}_- \sim -5.64071$ as shown in Eq.~(\ref{eq:dRN1}) and Eq.~(\ref{eq:dRN4}), respectively, in our method. 
We note that we have approximately read $r_0 \rightarrow r_\mathrm{m}+0$ in Eqs.~(\ref{eq:Eiroa3}) and (\ref{eq:Eiroa4}) as $r_0=r_\mathrm{m}+10^{-7}M$ for numerical calculations 
to get the values of $\bar{c}_+$ and $\bar{d}_+$ 
and we have used $r_0=r_\mathrm{m}-10^{-5}M$ for numerical calculations of $r_0 \rightarrow r_\mathrm{m}-0$ in Eqs.~(\ref{eq:Eiroa3}) and (\ref{eq:Eiroa4}) 
to get the values of $\bar{c}_-$ and $\bar{d}_-$. 
One would reduce numerical errors by calculating the integrals for the deflection angles more precisely. 

If the relation $\bar{d}_-=\bar{d}_+$ is also correct in other spacetimes, the relation would
be used to check our numerical method. 
In the Hayward spacetime with the marginally unstable photon sphere,
we have obtained 
$\bar{d}_+  \sim -5.60958$ and $\bar{d}_-  \sim -5.67374$ 
in our method and they imply that the relation $\bar{d}_-=\bar{d}_+$ is almost held.
We have approximately read $r_0 \rightarrow r_\mathrm{m}+0$ in Eqs.~(\ref{eq:Eiroa3}) and (\ref{eq:Eiroa4}) as 
$r_0=r_\mathrm{m}+3\times 10^{-5}M$
for numerical calculations 
to get the values of $\bar{c}_+$ and $\bar{d}_+$
and we have used 
$r_0=r_\mathrm{m}-2.4\times 10^{-5}M$ 
for numerical calculations of $r_0 \rightarrow r_\mathrm{m}-0$ in Eqs.~(\ref{eq:Eiroa3}) and (\ref{eq:Eiroa4}) 
to get the values of $\bar{c}_-$ and $\bar{d}_-$. 

In the Hayward spacetime with the marginally unstable photon sphere, 
we have obtained 
$\bar{c}_{+} \sim 6.01324$ 
in our numerical method and its difference from the value of $\bar{c}_{+} \sim 6.01316$ in Chiba and Kimura \cite{Chiba:2017nml} is tiny
but we have found that the value $\bar{c}_{+} \sim 4.95196$ obtained by the semianalytic calculation in Ref.~\cite{Tsukamoto:2020iez} should be modified.

Thus, we conclude that our method gives the correct values of $\bar{c}_{\pm}$ and $\bar{d}_{\pm}$ and that the semianalytic method considered 
by the author previously in Ref.~\cite{Tsukamoto:2020iez} gives the correct value of $\bar{d}_{+}$ but not the vaild value of $\bar{c}_{+}$.
We will revisit to modify the semianalytic calculation to obtain $\bar{c}_{\pm}$ correctly in forthcoming work~\cite{Igata:2026ivq}.

On this paper, we have concentrated on the extension of Eiroa, Romero, and Torres's method to gravitational lensing of the rays inside and outside of the marginally unstable photon sphere.
One may expect that we can apply it to the deflection angles~(\ref{eq:def0}) in the strong deflection limits $b \rightarrow b_\mathrm{m} \pm 0$ in spacetimes with a photon sphere and an antiphoton sphere. 
As shown in Refs.~\cite{Shaikh:2019itn,Tsukamoto:2021fsz}, however, $r_0$ does not approach $r_\mathrm{m}$ but another value in the limit $b \rightarrow b_\mathrm{m} - 0$.
Thus, the strong deflection limit $r_0 \rightarrow r_\mathrm{m} -0$ in Eqs.~(\ref{eq:Eiroa1})-(\ref{eq:Eiroa4}) is invalid and we need another extension of Eiroa, Romero, and Torres's method for the deflection angles~(\ref{eq:def0}) in the strong deflection limit $b \rightarrow b_\mathrm{m} - 0$ and it is left as a future work.

\section{Discussion}

The observed ring images of the supermassive black hole candidates at the centers of our Galaxy and the elliptic galaxy M87 by the Event Horizon Telescope~\cite{Akiyama:2019cqa,EventHorizonTelescope:2022wkp} 
are consistent with lensing images formed by synchrotron radiations from a hot plasma near the black holes according to their ray-traced general-relativistic magnetohydrodynamic simulations but the exotic compact objects such as naked singularity spacetimes, regular compact objects, and wormholes are not excluded~\cite{Akiyama:2019cqa,EventHorizonTelescope:2022wkp}.
This is because there are too many parameter regions for the exotic compact objects~\cite{Akiyama:2019cqa,EventHorizonTelescope:2022wkp} and astrophisical models~\cite{Gralla:2020pra}
to exclude the exotic cases.

We comment on constraints on metrics from the shadow observations of supermassive black hole candidates 
at the centers of our Galaxy and M87 by the Event Horizon Telescope~\cite{Akiyama:2019cqa,EventHorizonTelescope:2022wkp}.  
Event Horizon Telescope Collaborations~\cite{EventHorizonTelescope:2021dqv,EventHorizonTelescope:2022xqj}
constrained the parameter of the metrics of spacetimes from the observations and the sizes of the photon spheres under assumptions for simplify.
Thus, the constraints, which depend on the assumptions are not very reliable~\cite{Gralla:2020pra,Tsukamoto:2024gkz} and 
the constraints on the metrics should be confirmed by near-future space observations~\cite{Lupsasca:2024xhq,Johnson:2024ttr}.   

Event Horizon Telescope Collaborations do not detect the lensed images, formed by gravitational lensed rays near a photon sphere, separating from direct emissions of light sources, due to lack of angular resolution. 
To test metrics in spacetimes or exotic compact objects more accurately, we need to detect rays lensed in the strong gravitational field, separating from direct emissions of light sources  
since the lensed rays strongly depend on the metrics but do not depend on astrophysical environments much.
The lensed ring images formed by the rays bent near the photon spheres could be detected by the near-future space observations with lesser than $10$~$\mu$as of angular resolution~\cite{Lupsasca:2024xhq,Johnson:2024ttr}.  

A compact object with a marginally unstable photon sphere can be interesting exotic cases since it can be one of the most different cases from a Schwarzschild black hole among compact objects with diverging deflection angles. 
For the comparisons of the observable, we have to treat the rays inside of the marginally unstable photon sphere also since they can be brighter than rays nearly outside the marginally unstable photon sphere. 
In Refs.~\cite{Tsukamoto:2020iez} and \cite{Chiba:2017nml}, they do not give formulas for inside of the marginally unstable photon sphere while we and Ref.~\cite{Sasaki:2025web} do that.      
In Ref.~\cite{Sasaki:2025web}, we can see a complete and reliable treatment in the Reissner-Nordstr\"{o}m spacetime case.   
Our approach in this paper will give larger errors than Ref.~\cite{Sasaki:2025web} but it has a merit that it is easy to apply to other spacetimes.  
The precise values in Ref~\cite{Sasaki:2025web} in the Reissner-Nordstr\"{o}m spacetime case will be a milestone to improve our method more.  
We, however, emphasize that the larger errors in our numerical values in Table~I do not affect our discussion as we see below.

Supermassive black hole candidates at the centers of our Galaxy and M87 can be the most promising candidates to observe, specifically their gravitational lensing effects in strong gravitational fields since lensing images in their strong gravitational fields would be the largest seen from us.
In Table~I, as examples, the observable with the distances $d_{\mathrm{OS}}=16$kpc and $d_{\mathrm{OL}}=d_{\mathrm{LS}}=8$kpc, the mass $M=4\times 10^6 M_{\odot}$
in the Reissner-Nordstr\"{o}m spacetime and the Hayward spacetime is shown for the winding numbers $n=1$, $2$, and $3$.
As a reference, we show, in the Schwarzschild black hole case,
$2\theta_{\infty}=51.58$~$\mu$as,
$2\theta_{\mathrm{E}1+}=51.65$~$\mu$as, 
$\bar{s}_+ = 0.03$~$\mu$as,
$\mu_{1\mathrm{tot}+} (\phi) =1.6 \times 10^{-17}$,
with the source angle $\phi=1$ arcsecond,
and $\bar{r}_+ = 535$. 
For calculations for the Schwarzschild black hole, see, for examples, Ref. \cite{Tsukamoto:2021apr}. 

We will assume $10$~$\mu$as of the angular resolution which can be reached in near-future space observations~\cite{Lupsasca:2024xhq,Johnson:2024ttr}.
The angular resolution implies that we see the difference $\bar{s}_-$ but not $\bar{s}_+$ in principle.
In the usual lens configuration, from $\mu_{n \mathrm{tot} \pm}$ in Table~I, we notice that we would not detect the images with the winding numbers $n \geq 1$ since they are faint 
and the bright accretion flow falling into the central supermassive object will interfere with the detection 
even thought we have the angular resolution.  

In the rest of this section, we assume that we may apply the differences of the radii and the ratios of the magnifications of images with the winding numbers $n \geq 1$ in Table~I 
to the shadow observations on the supermassive black hole candidates at the centers of our Galaxy and M87.  
One may be afraid that our assumptions and discussion are invalid for the shadow observations 
since a usual lens configuration in which the light source is far away from the lens object cannot be suitable 
and a retrolensing configuration might be preferred but we might justify the differences of the lens configurations and of astrophysical situations 
as long as we concentrate on discussing only aspects which strongly depend on only the metric near the marginally photon sphere 
but do not depend on astrophysical environments and the lens configurations.

For the Schwarzschild spacetime, 
we would detect a ring image as a set of rays with 
indistinguishable diameters
$2\theta_{\mathrm{E}1+} \sim 2\theta_{\mathrm{E}\tilde{n}+} \sim 52$~$\mu$as,
where $\tilde{n}$ is an integer which is larger than $1$,
separating from another ring image of direct emissions of the bright accretion flow falling into the supermassive object.
From the value of $\bar{r}_+ = 535$, we notice that the contribution of the rays with winding numbers $n \geq 2$ to the ring is very small.

In the case of the Reissner-Nordstr\"{o}m spacetime with the marginally unstable photon sphere,
from the values of $\bar{s}_-=-15.187$~$\mu$as, $\bar{r}_-=2.95597$, $\bar{r}_+=17.3389$, and $\mu_{1\mathrm{tot}-}/\mu_{1\mathrm{tot}+}=4.2475$, 
we may detect two sets of images separating from the other ring image formed by the direct emissions; 
the set of outer ring image have indistinguishable diameters $2\theta_{\mathrm{E}1+} \sim 2\theta_{\mathrm{E}\tilde{n}+} \sim 2\theta_{\mathrm{E}\tilde{n}-} \sim 34$~$\mu$as 
and the inner image has a diameter $2\theta_{\mathrm{E}1-} \sim 6$~$\mu$as.
The magnification of the inner image is larger than the sum of the outer images by $1.7$ times.

For the Hayward spacetime with the marginally unstable photon sphere,
from $\bar{s}_-=-10.1007$~$\mu$as, $\bar{r}_-=9.60632$, $\bar{r}_+=17.0324$, and $\mu_{1\mathrm{tot}-}/\mu_{1\mathrm{tot}+}=14.403$, 
we may detect the set of outer ring image with indistinguishable diameters $2\theta_{\mathrm{E}1+} \sim 2\theta_{\mathrm{E}\tilde{n}+} \sim 2\theta_{\mathrm{E}\tilde{n}-} \sim 45$~$\mu$as, 
and an inner ring image with a diameter $2\theta_{\mathrm{E}1-} \sim 26$~$\mu$as,
separating from a ring image of the direct emissions.
The magnification of the inner ring image is greater than the sum of the outer ring images by $5.6$ times.
\begin{table*}[htbp]
 \label{table:I}
 \caption{The coefficients $\bar{c}_\pm$ and the terms $\bar{d}_\pm$ in the deflection angles of the rays deflected inside and outside of the marginally unstable photon spheres in the strong deflection limits~$b \rightarrow b_{\mathrm{m}} \pm 0$ in the Reissner-Nordstr\"{o}m (RN) spacetime obtained in our numerical method, by Sasaki \cite{Sasaki:2025web}, and a semianalytical method by Tsukamoto~\cite{Tsukamoto:2020iez} 
 and the ones in the Hayward spacetime in our method, by Chiba and Kimura~\cite{Chiba:2017nml}, and in the semianalytical method by Tsukamoto~\cite{Tsukamoto:2020iez}, 
 and observable calculated by them are shown in the case with distances $d_{\mathrm{OS}}=16$kpc and $d_{\mathrm{OL}}=d_{\mathrm{LS}}=8$kpc, and the mass $M=4\times 10^6 M_{\odot}$. 
 The reduced critical impact parameter $b_\mathrm{m}/M$, the diameter of the marginally unstable photon sphere $2\theta_{\infty}$,    
 the diameters of the Einstein rings $2\theta_{\mathrm{E}1 \pm }$, $2\theta_{\mathrm{E}2 \pm }$, and $2\theta_{\mathrm{E}3 \pm}$ with the winding number $n=1$, $2$, and $3$, respectively,  
 the difference of the radii $\bar{s}_\pm=\theta_{\mathrm{E}1 \pm }-\theta_\infty$, 
 the total magnifications of the pair images $\mu_{1\mathrm{tot}\pm} (\phi)$,  $\mu_{2\mathrm{tot}\pm} (\phi)$, and $\mu_{3\mathrm{tot} \pm} (\phi)$ with the source angle $\phi=1$ arcsecond, and with $n=1$, $2$, and $3$, respectively, and the ratio of the magnifications~$\bar{r}_\pm$ are shown.
 }
\begin{center}
\begin{tabular}{c ||c| c| c ||c |c |c} \hline
                                            &RN in our method &RN in \cite{Sasaki:2025web} &RN in \cite{Tsukamoto:2020iez} &Hayward in our method &Hayward in \cite{Chiba:2017nml}  &Hayward in \cite{Tsukamoto:2020iez}  \\ \hline
 $b_\mathrm{m}/M$                           &$3\sqrt{6}/2$    &$3\sqrt{6}/2$               &$3\sqrt{6}/2$                  &$25\sqrt{5}/12$       &$25\sqrt{5}/12$                  &$25\sqrt{5}/12$   \\
 $2\theta_{\infty}$ [$\mu$as]               &36.4727          &36.4727                     &36.4727                        &46.2429               &46.2429                          &46.2429           \\  \hline                  
 $\bar{c}_+$                                &6.67748          &6.67748                     &5.49892                        &6.01324               &6.01316                          &4.95196           \\ 
 $\bar{d}_+$                                &-5.59011         &-5.59108                    &-5.59108                       &-5.60958              &-                                &-5.62607          \\ 
 $2\theta_{\mathrm{E}1+}$ [$\mu$as]         &37.6267          &37.6262                     &36.8324                        &47.0155               &-                                &46.4819           \\ 
 $2\theta_{\mathrm{E}2+}$ [$\mu$as]         &36.5629          &36.5629                     &36.5008                        &46.3035               &-                                &46.2617           \\ 
 $2\theta_{\mathrm{E}3+}$ [$\mu$as]         &36.4879          &36.4879                     &36.4774                        &46.2531               &-                                &46.2460           \\ 
 $\bar{s}_+$ [$\mu$as]                      &0.57701          &0.57673                     &0.17987                        &0.38634               &-                                &0.11950           \\ 
 $\mu_{1\mathrm{tot}+} (\phi) \times 10^{19}$ &1063.82        &1063.19                     &324.595                        &888.554               &-                                &271.347           \\ 
 $\mu_{2\mathrm{tot}+} (\phi) \times 10^{19}$ &52.8675        &52.8477                     &16.4542                        &44.9320                &-                                &13.9129           \\ 
 $\mu_{3\mathrm{tot}+} (\phi) \times 10^{19}$ &6.58948        &6.58765                     &2.05397                        &5.61662               &-                                &1.7433            \\ 
 $\bar{r}_+$		                    &17.3389          &17.3350                     &16.9947                        &17.0324               &-                                &16.7917           \\ \hline
 $\bar{c}_-$                                &11.5658          &11.5657                     &-                              &10.4154               &-                                &-                 \\ 
 $\bar{d}_-$                                &-5.64071         &-5.59108                    &-                              &-5.67374              &-                                &-                 \\ 
 $2\theta_{\mathrm{E}1-}$ [$\mu$as]         &6.09793          &5.32981                     &-                              &26.0414               &-                                &-                 \\ 
 $2\theta_{\mathrm{E}2-}$ [$\mu$as]         &34.0762          &34.0367                     &-                              &44.6399               &-                                &-                 \\ 
 $2\theta_{\mathrm{E}3-}$ [$\mu$as]         &36.0681          &36.0631                     &-                              &45.9715               &-                                &-                 \\ 
 $\bar{s}_-$ [$\mu$as]                      &-15.187          &-15.571                     &-                              &-10.1007               &-                                &-                 \\ 
 $\mu_{1\mathrm{tot}-} (\phi) \times 10^{19}$ &4518.59        &4066.21                     &-                              &12798.4               &-                                &-                 \\ 
 $\mu_{2\mathrm{tot}-} (\phi) \times 10^{19}$ &1304.72        &1328.28                     &-                              &1141.19               &-                                &-                 \\ 
 $\mu_{3\mathrm{tot}-} (\phi) \times 10^{19}$ &173.347        &175.793                     &-                              &148.000             &-                                &-                 \\ 
 $\bar{r}_-$		                    &2.95597          &2.61464                     &-                              &9.60632               &-                                &-                 \\ \hline
\end{tabular} 
\end{center}
\end{table*}

\section*{Acknowledgements}
The author is grateful to T.~Igata, T.~Sasaki, H.~Asada, and an anonymous referee for their useful and valuable comments.
The author also would like to thank to M.~Saijo for his comments in the very early stage of this work.

\end{document}